# An Evolutional Algorithm for Automatic 2D Layer Segmentation in Laser-aided Additive Manufacturing

N. Liu[a], K. Ren[b*], W. Zhang[a], Y.F. Zhang[a], Y.X. Chew[b*], J.Y.H. Fuh[a], G.J. Bi[b]

[a] Department of Mechanical Engineering, National University of Singapore, Singapore
[b] Joining and Machining Group, Singapore Institute of Manufacturing Technology, 73 Nanyang Drive, 637662, Singapore

**Abstract**

Toolpath planning is an important task in laser aided additive manufacturing (LAAM) and other direct energy deposition (DED) processes. The deposition toolpaths for complex geometries with slender structures can be further optimized by partitioning the sliced 2D layers into sub-regions, and enable the design of appropriate infill toolpaths for different sub-regions. However, reported approaches for 2D layer segmentation generally require manual operations that are tedious and time-consuming. To increase segmentation efficiency, this paper proposes an autonomous approach based on evolutional computation for 2D layer segmentation. The algorithm works in an identify-and-segment manner. Specifically, the largest quasi-quadrilateral is identified and segmented from the target layer iteratively. Results from case studies have validated the effectiveness and efficacy of the developed algorithm. To further improve its performance, a roughing-finishing strategy is proposed. Via multi-processing, the strategy can remarkably increase the solution variety without affecting solution quality and search time, thus providing great application potential in LAAM toolpath planning. To the best of the authors' knowledge, this work is the first to address automatic 2D layer segmentation problem in LAAM process. Therefore, it may be a valuable supplement to the state of the art in this area.

## 1. Introduction

Toolpath planning for each sliced 2D layer is one of the essential tasks in laser aided additive manufacturing (LAAM) and other direct energy deposition (DED) processes , in which the scanning paths of the additive manufacturing tools, such as laser head and arc torch, are specified to ensure that the layer can be fully deposited [1,2]. There exist different types of path patterns, such as raster, zigzag, contour, etc [3]. Conventional toolpath planning approaches generally take the whole layer as input, and map it with a single deposition path pattern. However, this may cause various problems in actual part building [4]. In order to illustrate such problems, a simple example is given in Fig. 1, where Figs. 1a shows toolpath of the single raster pattern. It can be seen that *short path lengths* appear in Fig. 1a due to a sharp turn of the layer, which could easily lead to issues like inadequate powder melting, poor deposition quality/accuracy during the part building. To avoid such problem, an effective solution is to segment the original 2D layer into multiple sub-regions, and then fill each sub-region with appropriate toolpaths. In this way, the homogeneity of the generated toolpaths can be significantly improved (see Fig. 1b).

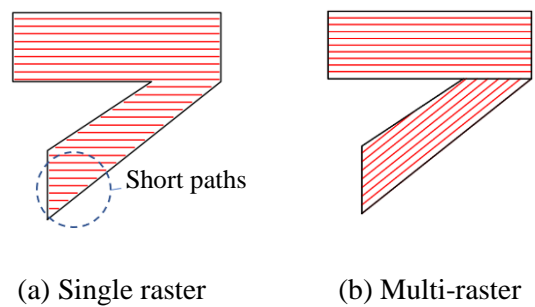

(a) Single raster     (b) Multi-raster

Fig. 1. Deposition paths with different strategies, (a) single raster path; (b) multiple regions - raster path.

The 2D layer segmentation problem shares certain similarities with 2D shape decomposition problems in the research area of computer vision (CV) [5–7], but clearly differentiates from the latter in terms of objectives. In CV, a 2D shape is decomposed for feature recognition and visual perception, while in LAAM, a 2D layer is segmented to facilitate deposition path planning towards better process control. As such, the existing CV methods cannot be applied directly in LAAM. In additive manufacturing (AM), this problem has drawn certain research attention over the years, due to its paramount importance in deposition quality control. Noting that traditionally generated deposition path may lead to voids, Ren et al. proposed to divide a complicated region into simpler sub-regions and treat these sub-regions separately for deposition path generation [8]. In their work, cell decomposition algorithm [9,10] was used for shape decomposition. However, since the cell decomposition algorithm was originally developed for robot coverage toolpath planning problem, the generated deposition path may contain a considerable amount of undesired short deposition paths. To generate void-free deposition path, Dong et al. proposed a medial axis-based approach by offsetting the medial axis towards its boundary [1,11]. Nonetheless, to execute the generated toolpath in actual LAAM, the deposition width needs to be altered continuously, which poses high operation cost. In addition, as pointed out by Lim et al., the medial axis-based approaches may lead to non-optimal configurations for regions of concavities [12]. Recently, Zhai et al. decomposed the 2D layer of a porous model into a set of topologically simple shapes via a generalized Voronoi diagram, and then generated a continuous toolpath [13]. The proposed divide-and-conquer approach could address a wide range of porous geometries. However, the resulting shapes could be too complicated for effective toolpath control in LAAM. To achieve effective deposition control, Michel et al. proposed a modular toolpath planning approach [4]. In their work, the original 2D layer was segmented into narrow rectangular shapes for individual toolpath generation. Their approach was successfully validated by metal deposition experiment, and comparative study revealed that the results were superior to other approaches. The main limitation of their 2D layer segmentation method is the interactive working manner, which heavily relies on user's input and decision. The entire process could be extremely time-consuming, thus severely hindering their application in LAAM. Another piece of related work is presented by Lim et al. towards automatic blocking of 2D shapes [12]. An evolutionary algorithm was employed to automatically segment the input shape into quadrilateral meshes for computational fluid dynamics (CFD) simulation.

As is stated above, conventional 2D layer segmentation in LAAM requires human intervention, which can be an extremely tedious and time-consuming task since the segmentation process needs to be performed for each 2D layer. Therefore, an automatic approach is urgently needed to segment a given layer into a set of basic geometries (called sub-regions). To this end, this paper aims to segment a 2D layer into a set of quasi-quadrilaterals automatically to enhance subsequent LAAM toolpath planning for improving process control (e.g., deposition dimensional accuracy, toolpath strategy selection, etc). Quasi-quadrilaterals are preferred sub-region shapes due to intuitive parameter control and relatively mature process simulation [14,15]. To achieve this, the key idea of this work is called identify-and-segment, which is to find the largest quasi-quadrilateral shape and segment it from the target layer. The remaining of the target layer is updated and the identify-and-segment process continues, until the whole target layer is completely segmented. For implementation, an evolutionary algorithm, i.e., genetic algorithm (GA), has been developed. To the best of the authors' knowledge, this work is the first to address the automatic 2D layer segmentation problem in LAAM process. It can also be extended to other DED processes.

The paper is organized in the following manner. Section presents an overview of the automatic 2D layer segmentation approach. Subsequently, the GA is implemented in Section 3, and the results are discussed in Section 4. Finally, Section 5 concludes the paper.

## 2. Overview of the automatic 2D layer segmentation approach

In AM, the slicing results of a designed part model is a set of 2D layers along the building direction. Each layer contains a set of closed loops (i.e., polygons). The objective of this work is to automatically segment the layer into a set of well-shaped quasi-quadrilaterals such that short infill path lengths and sharp turns can be effectively avoided. To achieve this, the overall idea is to convert the original problem to an iterative optimization problem aiming at finding the largest quasi-quadrilateral from the current shape at each iteration. The proposed approach works in an identify-and-segment manner. Specifically, the overall working mechanism is described as follows:

**Step 1.** Divide the original layer into a set of basic elements using the loop edges.

**Step 2.** Find the largest quasi-quadrilateral by merging the existing basic elements.

**Step 3.** Update the existing basic elements by deleting the elements in the quasi-quadrilateral obtained from Step 2.

**Step 4.** Repeat Steps 2-3 until no basic element remains.

Step 1 is called the pre-processing of the input layer, while Steps 2-4 specify the proposed identify-and-segment process. The details of these processing steps are further illustrated in the following sections.

3.1. Pre-processing

Given a 2D layer consisting of a set of polygons, each represented by a sorted point set $P = \{p_1, p_2, \ldots, p_m\}$, the first step is to identify the *deposition area* or the material area. A polygon that does not require infill deposition is defined as a *hole loop*, while a polygon to be fully/partially infilled is defined as a *material loop*. After the identification of all hole loops and material loops, the deposition area can then be determined as the area between material loops and hole loops, if any. A simple example is given in Fig. 2a, whereby the blue zone is identified as the deposition area.

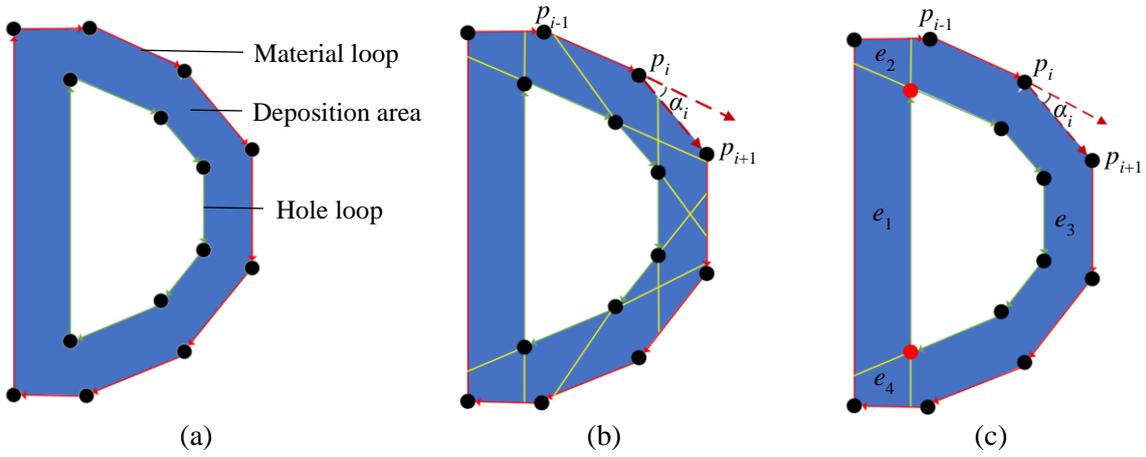

Fig. 2. An example showing the pre-processing of a simple 2D layer, (a) deposition area identification; (b) decomposition at each vertex; (c) decomposition at sharp turns ($\alpha_{max}=30°$).

In the second step, the deposition area is divided. An intuitive idea is to decompose the original polygon at each vertex by extending the edges to divide the deposition area, as shown in Fig. 2b. However, this may result in too many redundant elements, causing a combinatorial-explosive problem for the subsequent quasi-quadrilateral searching process. Since sharp turns in the resulted sub-regions should be avoided as much as possible, we will only perform the division at sharp turns. Without loss of generality, the *local angle* $\alpha_i$ at vertex $p_i$ is defined as the angle between vectors $\overrightarrow{p_{i-1}p_i}$ and $\overrightarrow{p_ip_{i+1}}$, as shown in Fig. 2b. If a local angle is larger than a threshold value ($\alpha_{max}$), it is defined as a sharp turn. Subsequently, only those edges at sharp turns will be extended to divide the deposition area into a set of elements $E = \{e_1, e_2, \ldots, e_n\}$, which are considered as *basic elements*. Through trial and error, decent results can be obtained when $\alpha_{max}$ is set between 20° and 30°. For the example shown in Fig. 2a, using $\alpha_{max} = 30°$, the local angle at each vertex is checked and only two vertices are identified as sharp turns (see Fig. 2c). This layer is then divided by extending the edges at these two sharp turns and four basic elements are obtained as shown in Fig. 2c. With this heuristic, the number of basic elements is significantly reduced compared with the exhaustive dividing results in Fig. 2b.

3.2. Identify-and-segment

The output of pre-processing is a basic element set $E$. Very often, the elements in the set $E$ are not suitable for path planning directly as they will lead to excessive short paths due to their limited sizes. To solve this problem, this paper utilizes an identify-and-segment process. Specifically, the largest quasi-quadrilateral obtained by merging the basic elements is firstly identified. Subsequently, the obtained quasi-quadrilateral is segmented from the layer and the basic element set $E$ is updated accordingly. For illustration, an output of the identification process for the layer in Fig. 2a is presented in Fig. 3a, where the shaded rectangle on the left is identified by merging elements $e_1$, $e_2$ and $e_4$ in Fig. 2c. This rectangle is then stored in the sub-region list, and the current basic element set $E$ is updated by deleting the merged basic elements ($e_1$, $e_2$ and $e_4$), as shown in Fig. 3b. To completely segment the layer, the identify-and-segment process continues until there is no element left in the basic element set $E$. In this particular case, $e_3$ is the only element left and will form the second sub-region.

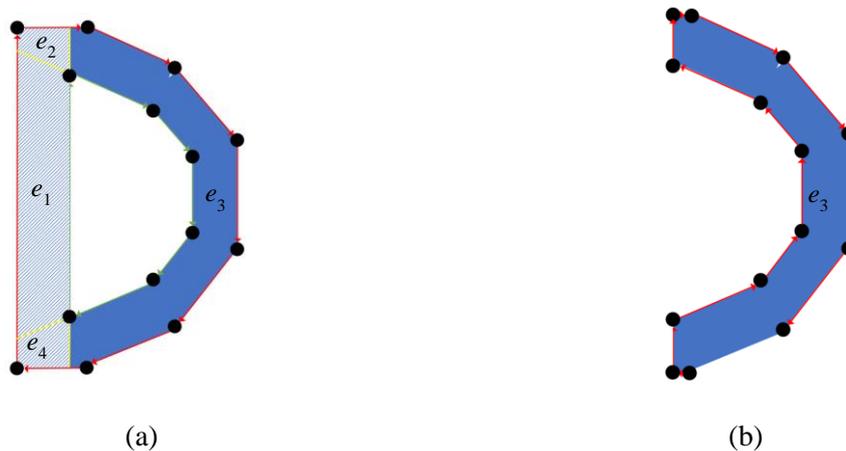

(a)          (b)

Fig. 3. Example of identify-and-segment, (a) identification of the largest quasi-quadrilateral; (b) updated basic element set

## 3. Search for the largest quasi-quadrilateral – an evolutional algorithm

In the proposed identify-and-segment approach, the largest quasi-quadrilateral needs to be identified by merging the existing basic elements. Apparently, an exclusive search method is not practical for such a combinatorial problem. In this work, an evolutional approach is employed due to its flexible customization and decent capability in solving various optimization problems [16]. A genetic algorithm (GA) has thus been developed and its overall procedure is briefly described as follows:

**Step 1. Population initialization.** Randomly initialize a fixed number (population size, $N_{PS}$) of chromosomes in the first generation.

**Step 2. Fitness evaluation.** Decode each chromosome in the current generation to its corresponding solution and calculate its fitness value according to a specially designed fitness evaluation function.

**Step 3. Stopping criterion check.** If the specified stopping criterion is satisfied, go to Step 5.

**Step 4. Reproduction.** Generate a new generation with the same number of chromosomes according to a specific reproduction mechanism and go to Step 2.

**Step 5.** Output the obtained solution and stop.

4.1. Chromosome representation and population initialization

In the developed GA, a chromosome ($S$) is essentially a moving rectangle (m-rectangle) encoded in a five-gene vector, $S = [w, h, t_x, t_y, \theta]$, where $w$ and $h$ are the width and height of the rectangle, respectively; $t_x$ and $t_y$ are the translation of the rectangle with reference to the origin along $x$-axis and $y$-axis, respectively; $\theta$ is the rotation angle in $x$-$y$ plane (see Fig. 4a). In GA implementation, a chromosome is firstly decoded into an m-rectangle and superimposed onto the existing basic elements in $E$. A corresponding solution can then be obtained by merging all basic elements intersecting with the m-rectangle. In this way, a mapping from a chromosome to its corresponding solution is successfully established, indicating the solution space can be traversed by varying chromosomes. An example is given in Fig. 4b, where the chromosome (m-rectangle) and its corresponding solution ($e_1$ and $e_4$) are shown.

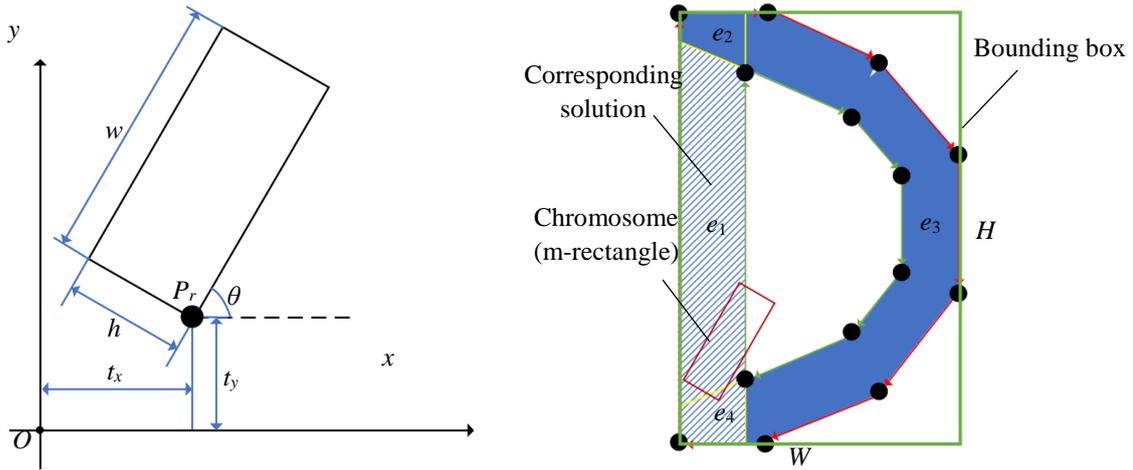

Fig. 4. Chromosome representation and decoding, (a) chromosome representation, (b) chromosome decoding and its corresponding solution.

The first step in the developed GA is to generate $N_{PS}$ chromosomes as the first generation. The procedure can be described as follows (see Fig. 4):

**Step 1.** Find the bounding box (width: $W$, height: $H$, see Fig. 4b) covering the current basic element set $E$.

**Step 2.** Randomly generate $N_{PS}$ chromosomes (m-rectangles) inside the bounding box. To generate a single m-rectangle, a reference point (see $P_r$ in Fig. 4a) is randomly generated within the bounding box ($t_x$ and $t_y$ are defined). Subsequently, $w$, $h$, and $\theta$ are randomly generated satisfying $w < W$, $h < H$, and $0 < \theta < 180°$.

**Step 3.** Check the validity of each chromosome. If the corresponding solution of a chromosome does not intersect with any basic element, it is invalid. If all chromosomes are valid, go to Step 5.

**Step 4.** For each invalid chromosome, replace it with a re-generated one within the bounding box, and go to Step 3.

**Step 5.** Output the first generation of chromosomes.

4.2. Fitness evaluation

In each generation, the chromosomes need to be evaluated to obtain their fitness values. A properly designed fitness evaluation function will effectively guide the search towards the optimal/near-optimal solution. As shown in Fig. 5, an m-rectangle intersects with the basic elements, leading to multiple intersection areas ($S_0$-$S_3$) in mm$^2$. The effects of these intersection areas on the quality of the resulted corresponding solution are discussed as follows:

- $S_0$ is the common area between the m-rectangle and the deposition area. Generally, a chromosome with a larger $S_0$ is preferred.
- $S_1$ is the area of the corresponding solution. Since the search is for the largest quasi-quadrilateral, a chromosome with a larger $S_1$ is preferred.
- $S_2$ represents the area of the m-rectangle outside the material loop. A very large $S_2$ will probably lead to an invalid chromosome/solution in the search process, thus slowing down the GA search process. Therefore, $S_2$ should be kept at low level to increase the search efficiency.
- $S_3$ denotes the area of m-rectangle inside the hole loop. Since segments containing holes inside will significantly complicate subsequent LAAM process (path planning, parameter control and simulation, etc.), a smaller $S_3$ is preferred.

One more factor affecting the quality of the corresponding solution is the number of the sharp turns ($N_{st}$) in the corresponding solution. Since the search is for quasi-quadrilateral, $N_{st}$ should be close to 4 as much as possible.

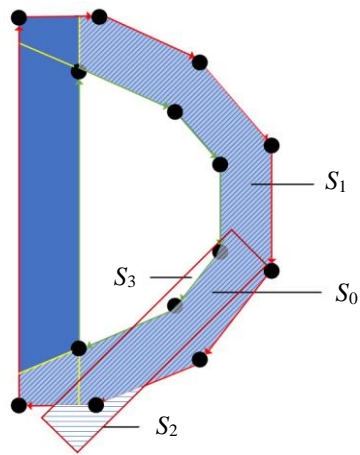

Fig. 5. Definition of $S_0, S_1, S_2, S_3$ in fitness function.

Based on the above heuristics, a generic fitness function is proposed as in Eq. (1).

$$F = -\frac{c_0 S_0 + c_1 S_1}{\exp(|4 - N_{st}|) + c_2 S_2 + c_3 S_3} \tag{1}$$

where the negative sign indicates a better chromosome has a smaller $F$. The values of the four coefficients are determined through trial and error. The final setting for the implemented GA is as follows: $c_0 = 0.001$; $c_1 = c_3 = 1$. If $S_2$ is larger than 50% of the m-rectangle area, $c_2 = 1$; otherwise, $c_2 = 0$.

4.4. Reproduction

After fitness evaluation for the current generation is completed, if the stopping criterion is not reached, the current generation will be replaced by a new generation through reproduction. For this reproduction process, $N_{new}$% chromosomes will be randomly selected from the current generation as parents for offspring generation, while the other (100-$N_{new}$)% chromosomes will be directly placed in the new generation. For the $N_{new}$% chromosomes, two kinds of GA operators, crossover or mutation, will be applied to generate offsprings. For a given parent chromosome, the mutation or crossover operator will be invoked by comparing a random number $X$ from (0, 1) and a predetermined probability $\rho$. Mutation will be activated if $X > \rho$; otherwise, the crossover will be applied. The procedure of a mutation operation is described as follows (see Fig. 6):

**Step 1.** Randomly select one chromosome from the $N_{new}$% chromosomes as the parent.
**Step 2.** Randomly select one or several gene positions.
**Step 3.** Replace the genes at selected positions with newly generated genes.
**Step 4.** Check the validity of the newly generated chromosome. If not valid, go to Step 3.
**Step 5.** Evaluate the fitness of the offspring. If it has a smaller fitness than its parent, place it in the new generation; otherwise, place the parent in the new generation.

|  | **Mutation** | **Crossover** | |
|---|---|---|---|
| **Parent** | $[w, \underline{h}, t_x, t_y, \underline{\theta}]$ | $[w, \underline{h}, \underline{t_x}, \underline{t_y}, \theta]$ | $[\underline{w^*}, \underline{h^*}, \underline{t_x^*}, \underline{t_y^*}, \underline{\theta^*}]$ |
| **Offspring** | $[w, h', t_x, t_y, \theta']$ | $[w^*, h, t_x^*, t_y^*, \theta]$ | |

Fig. 6. Example showing the mechanism of GA operators.

Similarly, a crossover operation is as follows (see Fig. 6):

**Step 1.** Randomly select two chromosomes from the $N_{new}$% chromosomes as the parents and initialize the offspring as the first parent.

**Step 2.** Randomly select one or several gene positions in the offspring.

**Step 3.** Replace the genes at selected positions with those in the second parent.

**Step 4.** Check the validity of the newly generated offspring. If not valid, go to Step 2.

**Step 5.** Evaluate the fitness of the offspring. If it has a smaller fitness than its first parent, place it in the new generation; otherwise, place its first parent in the new generation.

Upon the completion of the GA operation, a new generation consisting of $N_{PS}$ chromosomes is reproduced. This proposed reproduction scheme not only guarantees elitism automatically, but also reserves the variety of chromosomes.

For a given input 2D shape, the GA stops when the best fitness value remains unchanged for $N_s$ generations. The output of the GA is the chromosome with the best fitness value. Through trial and error, the implemented GA achieved decent performance with its parameters set as $[N_{PS}, N_{new}, \rho, N_s]$ = [40, 90, 0.5, 60].

## 4. Results and discussion

The developed automatic 2D layer segmentation algorithm has been implemented in Python on a Linux laptop (Intel Core i7-9750H CPU, 16GB RAM). This section firstly presents some case studies that were carried out to test the effectiveness and efficiency of the algorithm. Subsequently, a roughing-finishing strategy is proposed for further increase of its performance.

5.1. Case studies

In case study 1, a test part similar to that in [4] was chosen as a benchmark. The 3D part model is shown in Fig. 7a, which was firstly sliced using an in-house developed slicing algorithm as shown in Fig. 7b. One slice was randomly chosen as the input layer for the GA segmentation algorithm (see Fig. 7c).

The parameters were set as [$\alpha_{max}$, $N_{PS}$, $N_{new}$, $\rho$, $N_s$] = [30°, 40, 90, 0.5, 60]. The input layer was pre-processed into basic elements as shown in Fig. 7d. Then, the GA was run until the layer was fully segmented which took a total of 13 iterations (segmentation actions). The total running time is 75.72s. The obtained chromosomes and their corresponding solutions at each iteration are presented in Fig. 7e. It can be seen the obtained chromosomes could successfully merge the basic elements into quasi-quadrilateral shapes (e.g., straight/curved rectangles). The search history (best fitness for each generation) of the first iteration is shown in Fig. 8a in which the best solution is reached after 140 generations. Fig. 8b shows the time consumed for each iteration. It can be seen that the first iteration took the longest running time as the input shape is the most complex. The running time for the subsequent iterations decreased dramatically as the remaining yet-to-segment shape gets simpler. The final segmentation result is presented in Fig. 7f. This result is comparable to manual segmentation result in [4]. Compared with the time-consuming manual segmentation process, the time saving is very significant.

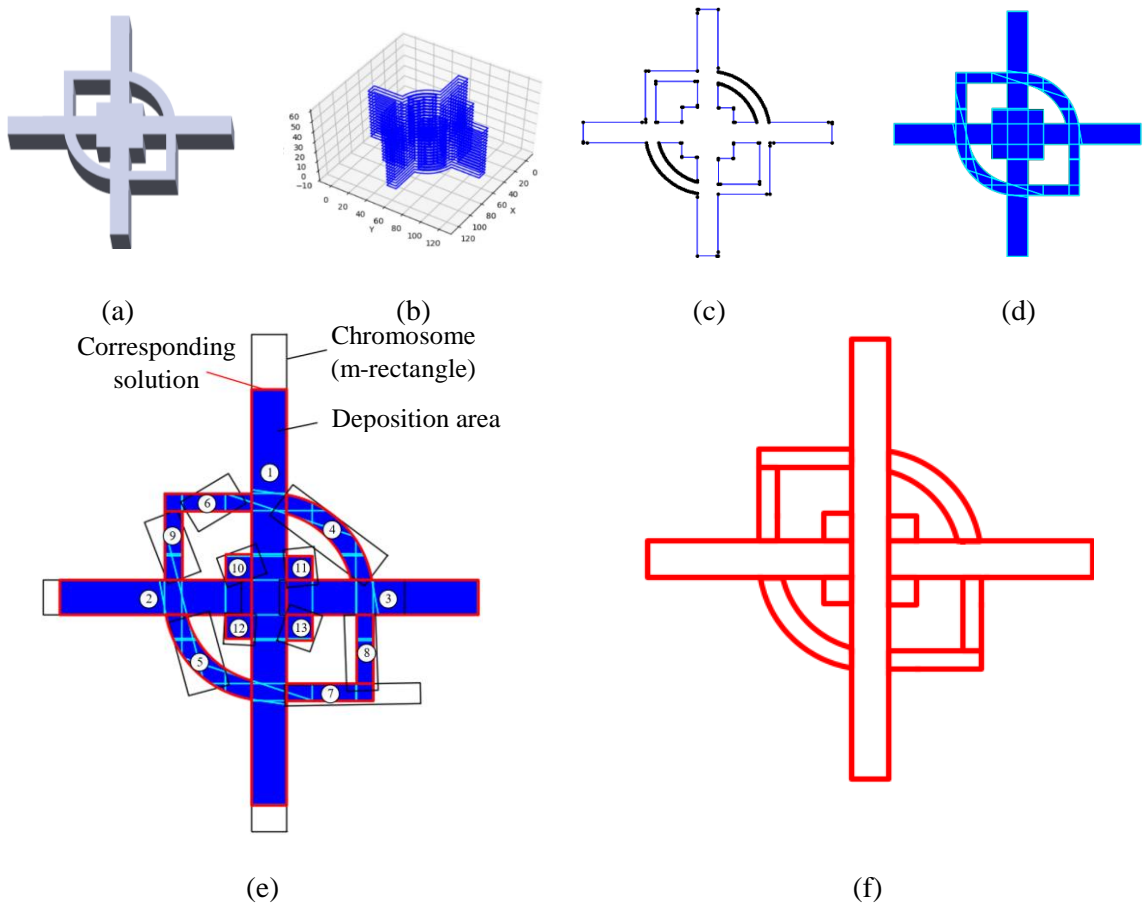

(a)     (b)     (c)     (d)

Corresponding solution    Chromosome (m-rectangle)    Deposition area

(e)     (f)

Fig. 7. Case study 1, (a) 3D part model; (b) slicing result; (c) input 2D layer; (d) pre-processing result; (e) iterative segmentation results; (f) segmentation result.

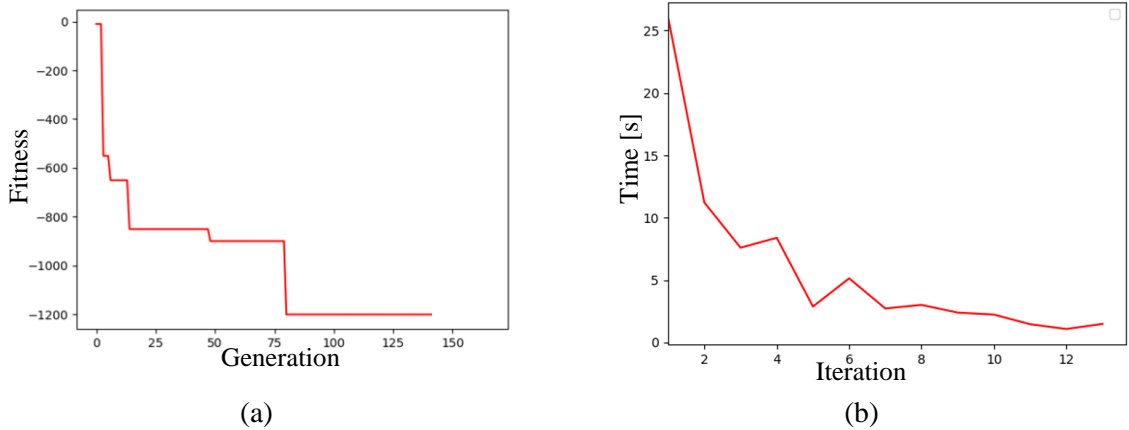

(a)                                                                                          (b)

Fig. 8. GA search history, (a) fitness evolution in 1st iteration; (b) running time in each iteration.

To study the generalization capability of the developed segmentation algorithm, another two parts shown in Fig. 9 were tested. Similarly, each 3D model was sliced first, and one slice was randomly chosen as the input layer for segmentation. The segmentation results for the two cases are shown in Figs. 9c and 9f, respectively. This clearly shows the decent generalization capability of the developed segmentation algorithm.

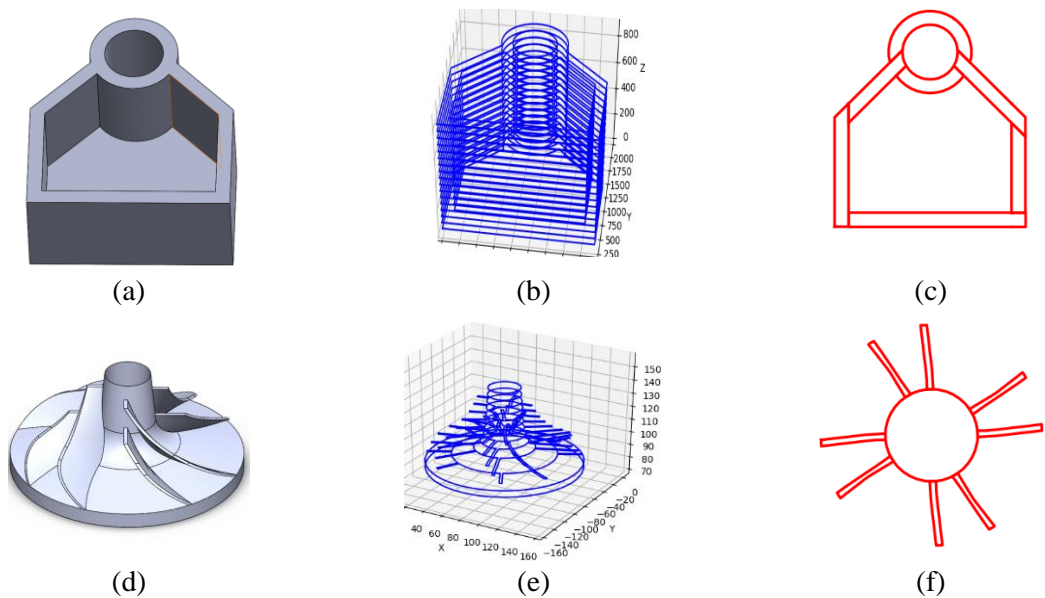

(a)                                          (b)                                          (c)

(d)                                          (e)                                          (f)

Fig. 9. Case studies of another two parts, (a)-(c) 3D model, slicing results, and segmentation result of a hinge; (d)-(f) 3D model, slicing results, and segmentation result of an impeller.

## 5.2. A roughing-finishing strategy

Although the efficiency of the developed segmentation algorithm is reasonably good, the performance was further improved by employing a roughing-finishing running strategy, which aims to improve the overall efficiency without compromising the final segmentation quality. Specifically, the developed segmentation algorithm is invoked in roughing stage and finishing stage. At roughing stage, the GA re-organizes the basic elements into roughly segmented shapes. Subsequently, at finishing stage, the GA treats the roughly segmented shapes as basic elements, and further re-organizes these input shapes into final segmentation results. This roughing-finishing strategy is implemented by adjusting the stopping criterion, $N_s$. In this study, the stopping criteria, $N_{s-rough}$ and $N_{s-finish}$ are set as 20 and 30 for roughing stage and finishing stage, respectively. For comparison, the same layer in Fig. 7c was taken as input. The segmentation results from both stages are shown in Fig. 10a and Fig. 10b, respectively. The total running times for the two stages are 22.77s and 15.29s, respectively. Compared with the single-mode running in Fig. 7, a time saving of about 50% is achieved (75.72s vs. 38.06s). The main reason is as follows. With a much relaxed stopping criterion ($N_{s-rough} = 20$), the roughing search stops quickly but prematurely at non-optimal solutions (see Fig. 10b). However, at the same time, the result from the roughing stage significantly reduces the total number of basic segments. In this way, the solution space at finishing stage is greatly reduced, and the non-optimal issue can be successfully resolved at finishing stage by using a slightly stricter stopping criterion ($N_{s-finish} = 30$). Fig. 10d shows the final segmentation result that is comparable to Fig. 7f. Therefore, it can be safely concluded that the segmentation efficiency can be further improved by the roughing-finishing strategy, while maintaining the segmentation quality.

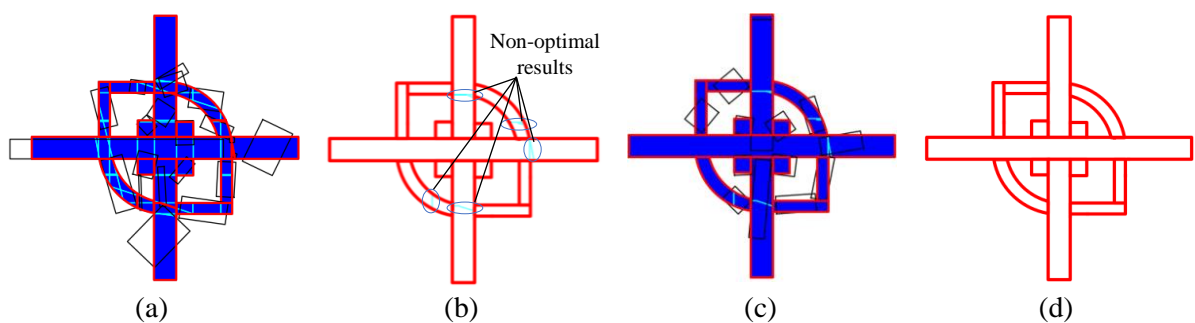

(a)     (b)     (c)     (d)

Fig. 10. Segmentation results using a roughing-finishing strategy, (a)-(b) iterative and final segmentation results at roughing stage; (c)-(d) iterative and final segmentation results at finishing stage.

Another interesting phenomenon was observed during the implementation of the roughing-finishing strategy. The relaxation of the stopping criterion $N_s$ leads to a significant increase in the solution variety. In addition to the segmentation results shown in Fig. 10c, other segmentation results could also be obtained, as shown in Fig. 11. The reason for this phenomenon is probably because the roughing stage prunes the original solution space in a nondirectional manner, leading to diversified solution space for the finishing stage. Nonetheless, the solution quality is also high judging by visual inspection. This feature is of high significance in AM software development as modern CPU usually has multiple cores and the same algorithm can be invoked simultaneously by multiprocessing. Without incurring significant amount of search time (50.81s in Fig 11 vs 38.06s in Fig. 10), more diversified segmentation results can be obtained simultaneously for users to choose. This particular capability of generating various feasible segmentation results is very useful if coupled with an efficient LAAM process simulation programme [14,15] for the toolpath planning.

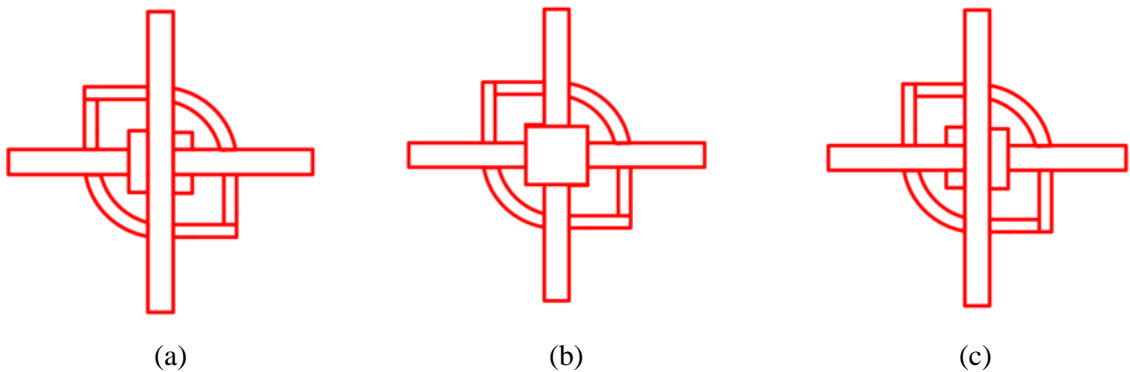

(a)　　　　　　　　　　(b)　　　　　　　　　　(c)

Fig. 11. Diversified segmentation results by 3-core multiprocessing, total search time: 61.27s.

## 5. Conclusions and future work

This paper presents a novel GA-based automatic 2D layer segmentation algorithm for LAAM process. Case studies have demonstrated both effectiveness and efficiency of the developed algorithm. Employing an iterative identify-and-segment manner, the developed algorithm could effectively segment the input layer into well-shaped quasi-quadrilaterals (e.g.., straight/curved rectangles). To further improve its performance, a roughing-finishing strategy has been proposed and implemented. Test results are very promising, with search time reduced and maintained segmentation quality. Besides, the diversity of the segmentation results is also increased, providing more candidates for the decision-maker. To the best of the authors' knowledge, the proposed approach is the first automatic 2D layer segmentation algorithm to address 2D layer segmentation problem in LAAM process. The segmented shapes could significantly benefit subsequent tasks such as toolpath planning, parameter control and simulation, thus providing great application potential in LAAM process. Noting that the proposed algorithm may be inevitably limited by its evolutionary nature (GA), the next stage of our work is to apply latest artificial intelligent techniques to address 2D shape segmentation algorithm. The influence of different 2D layer segmentation results on actual part building quality will be also investigated.

**Acknowledgements**

This research was supported by Agency for Science, Technology and Research (A*Star), Republic of Singapore, under the IAF-PP program "Integrated large format hybrid manufacturing using wire-fed and powder-blown technology for LAAM process", Grant No: A1893a0031.